# Optimization of Al/AlO$_x$/Al-Layer Systems for Josephson Junctions from a Microstructure Point of View

S. Fritz[1*], L. Radtke[2], R. Schneider[1], M. Weides[2,3] and D. Gerthsen[1*]

[1]Laboratory for Electron Microscopy, Karlsruhe Institute of Technology (KIT), 76131 Karlsruhe, Germany

[2]Physikalisches Institut, Karlsruhe Institute of Technology (KIT), 76131 Karlsruhe, Germany

[3]School of Engineering, University of Glasgow, G12 8LT, Glasgow, United Kingdom.

*Correspondence and requests for materials should be addressed to D.G. (email: dagmar.gerthsen@kit.edu) or S.F. (email: stefan.fritz@kit.edu)

**Abstract**

Al/AlO$_x$/Al-layer systems are frequently used for Josephson junction-based superconducting devices. Although much work has been devoted to the optimization of the superconducting properties of these devices, systematic studies on influence of deposition conditions combined with structural analyses on the nanoscale are rare up to now. We have focused on the optimization of the structural properties of Al/AlO$_x$/Al-layer systems deposited on Si(111) substrates with a particular focus on the thickness homogeneity of the AlO$_x$-tunnel barrier. A standard high-vacuum electron-beam deposition system was used and the effect of substrate pretreatment, different Al-deposition temperatures and Al-deposition rates was studied. Transmission electron microscopy was applied to analyze the structural properties of the Al/AlO$_x$/Al-layer systems to determine the thickness homogeneity of the AlO$_x$ layer, grain size distribution in the Al layers, Al-grain boundary types and the morphology of the Al/AlO$_x$ interface. We show that the structural properties of the lower Al layer are decisive for the structural quality of the whole Al/AlO$_x$/Al-layer system. Optimum conditions yield an epitaxial Al(111) layer on a Si(111) substrate with an Al-layer thickness variation of only ±1.6 nm over more than 10 μm and large lateral grain sizes up to 1 μm. Thickness fluctuations of the AlO$_x$-tunnel barrier are minimized on such an Al layer which is essential for the homogeneity of the tunnel current. Systematic variation of the Al-deposition rate and deposition temperature allows to develop an understanding of the growth mechanisms.

## I. Introduction

Superconducting devices are frequently based on Josephson junctions (JJ) fabricated on the basis of Al/AlO$_x$/Al-layer systems where a thin AlO$_x$ layer serves as tunnel barrier. JJs are used, e.g., in superconducting quantum bits for the realization of quantum information circuits [1], single photon detectors [2], radiation detectors [3], single electron transistors [4] and superconducting quantum interference devices in magnetometers [5, 6]. The structural properties of the layer system have a profound influence on the performance of superconducting devices and on noise that limits detection sensitivity and coherence. For example, thickness variations of the AlO$_x$-tunnel barrier is a critical problem because the tunnel current scales exponentially with tunnel barrier thickness. The homogeneity of JJs is particularly crucial for complex superconducting circuits for quantum information processing, which contain a large number of JJs. A previous study by Zeng *et al.* [7] has in this context shown that less than 10 % of the total AlO$_x$-tunnel barrier area in JJs is active in the tunnelling process in their Al/AlO$_x$/Al-based JJs due to thickness variations of the amorphous AlO$_x$ layer. This is disadvantageous with respect to performance and necessitates optimization of the thickness homogeneity of the tunnel barrier. AlO$_x$-thickness variations are predominantly caused by grain boundary grooving in the lower Al-electrode layer as shown by Nik *et al.* [8] and our group [9]. Hence, microstructure and homogeneity of the lower Al layer determine to a large degree the properties of the whole Al/AlO$_x$/Al-layer system and have to be optimized to provide the best possible surface for the formation of an AlO$_x$-tunnel barrier with homogeneous thickness. In fact, an Al layer grown epitaxially on a suitable substrate with an atomically flat surface would be ideal.

Epitaxial growth of Al on Si substrates has already been realized by ultrahigh vacuum (UHV)-based deposition techniques like molecular beam epitaxy (MBE) [10] or UHV evaporation [11, 12]. However, UHV deposition systems are elaborate to operate and in general not well suited for JJ fabrication because shadow evaporation techniques [13, 14] are difficult to implement. Up to now, high-vacuum (HV) electron-beam deposition systems, such as the *Plassys MEB 550S* system, are mainly used for JJ fabrication. With Al-deposition parameters, which are typically applied for JJ fabrication in HV systems (deposition rates of 0.1 – 1.2 nm/s and substrate temperatures between room temperature and 200 °C) [5, 7, 15–19], epitaxial growth of Al was not reported up to now.

Nevertheless, previous work in UHV systems give useful hints on prerequisites for optimizing Al deposition. Chemical substrate cleaning prior to Al deposition [20, 21] is the first step on the path to epitaxial Al growth. A clean Si/Al interface also improves the electrical properties of whole JJ [22] and is thus not only beneficial for Al growth. Al(111) surfaces have the lowest surface energy in Al [23] and are best suited for obtaining epitaxial Al layers with a homogeneous thickness. Even epitaxial growth of $\gamma$-$Al_2O_3$(111) on Al(111) has been observed under UHV conditions in a MBE system [24, 25] because $AlO_x$ layers on Al(111) have the lowest calculated critical thickness above which crystalline $\gamma$-$Al_2O_3$ layers are thermodynamically preferred over amorphous $AlO_x$ layers [24]. Despite the lattice mismatch of 25.5% between Al and Si, epitaxial growth of Al(111) can be best achieved on Si(111) substrates [26, 27]. Using Si(100) substrates, Al tends to grow in [110] direction [28], which is unwanted for the oxidation process [24]. Moreover, the low surface energy of Al(111) is promising for achieving Al layers with homogeneous thickness. We note that we will consider growth of Al(111) parallel to Si(111) as epitaxial growth, although grains can be rotated around the [111]-growth direction and the layer will therefore not be single-crystalline.

In this work the structural properties of Al/$AlO_x$/Al-layer systems deposited on Si(111) substrates were correlated with growth conditions. The structural properties were in detail investigated by transmission electron microscopy (TEM). Substrate pretreatment, substrate temperature during Al deposition and Al-deposition rate were systematically varied to optimize the structural quality of Al/$AlO_x$/Al-layer systems in a standard electron-beam deposition system (*Plassys MEB 550S*) with a base pressure in the HV range. In particular, AlO$_x$-tunnel barriers with homogeneous thickness were obtained by achieving epitaxial growth of the lower Al layer, which provides a road map to optimized JJ fabrication.

## II. Experimental techniques

Al/AlO$_x$/Al-layer systems were deposited on single-crystalline Si(111)-substrates in a *MEB 550S* (*PLASSYS Bestek, Marolles-en-Hurepoix, FR*) electron-beam physical vapor deposition system with a base pressure in the HV regime where a pressure of $5 \cdot 10^{-7}$ mbar is archived after 1h of pumping. Pure N$_2$ is used for venting and purging the chamber. The system is equipped with a *kaufman source*, which generates an Ar/O-Plasma (4 sccm Ar and 0.5 sccm O$_2$) with an acceleration voltage of 200 V and an ion current of 10 mA for removing carbonaceous contamination from the substrate.

In the first step, cleaning of the Si(111) substrates was optimized and the influence of different procedures was studied. All substrates were chemically treated to remove the protective resist layer by dipping the substrates successively in NEP (N-ethyl-2-pyrrolidon), isopropyl alcohol and deionized water. In some experiments, an additional HF-dip process was applied to remove the native silicon oxide (SiO$_x$) which remains after the first chemical cleaning. In this process the substrate is dipped in the buffered oxide etch *BOE 7:1* (12.5 % HF and 87.5 % NH$_4$F) (*Microchemicals GmbH, Ulm, Germany*) for 45 s. During the HF-dip etching, the SiO$_x$ layer is completely removed and an atomically flat hydrogen-terminated surface is formed [29]. The substrate is then rinsed with deionized water to remove the *BOE 7:1* and stop the etching process. Transfer and insertion of the HF-cleaned substrate in the *MEB 550S* system have to be completed in less than one minute to avoid re-oxidation in air. The load lock is pumped to $10^{-6}$ mbar and the molybdenum sample plate is heated by a resistance heating wire to 175 °C to desorb residual moisture from the substrate. After 25 min at 175 °C, the substrate temperature is increased to 700 °C for 20 min to thermally desorb hydrogen, fluorine and residual oxide [20]. According to McSkimming *et al.* [26], during this treatment the Si substrate forms a Si(111) 7x7 reconstructed surface which remains stable even at lower temperatures. We could not verify the Si(111) 7x7 surface reconstruction because a reflection high-energy electron diffraction system is not available in our deposition system, but values for Al-thickness variations for our epitaxially grown samples (cf. Table 2) are in agreement with values reported by McSkimming *et al.* [26] for 100 nm Al deposited on Si(111) 7x7 in their UHV system. Also, according to McSkimming, epitaxial Al films only occur on Si(111) 7x7 or Si(111) $\sqrt{3}$x$\sqrt{3}$ surfaces whereas unreconstructed Si(111) 1x1 surfaces lead to polycrystalline layers.

The lower Al layer is deposited by electron-beam evaporation from a pure Al target. Five samples with HF-dip and high-temperature treatment were fabricated with Al deposition at different substrate temperatures $T_s$ between 100 °C and 300 °C. This temperature range was chosen because is it relevant for forming $AlO_x$-tunnel barriers by oxidation of the Al surface and eventually even grow crystalline $AlO_x$ layers [24]. Growth of the layer system at room temperature, although frequently applied, is not compatible with the high-temperature step to generate a 7x7 reconstructed Si(111) surface because cooling to room temperature requires several hours and substrate holder cooling is not available in our deposition system. Contamination will occur during cooling to room temperature, which prevents epitaxial Al growth.

Al-deposition rates $r$ at $T_s$ = 100 °C were varied from 0.1 nm/s to 1 nm/s which are basically the limits of our deposition system. Substrate temperature and deposition rate have the strongest influence on the microstructure of the deposited layer and are used as sample denotations (cf. Table 1). Temperatures were controlled by a resistance temperature sensor on the backside of the sample plate and deposition rates were controlled by a piezoelectric sensor. The Al deposition was terminated at 100 nm layer thickness.

| sample | deposition temperature [ºC] | deposition rate [nm/s] |
|---|---|---|
| Al$_{300\_0.1}$ | 300 | 0.1 |
| Al$_{200\_0.1}$ | 200 | 0.1 |
| Al$_{100\_0.1}$ | 100 | 0.1 |
| Al$_{100\_0.5}$ | 100 | 0.5 |
| Al$_{100\_1.0}$ | 100 | 1.0 |

**Table 1**. Deposition conditions for the lower Al layer with corresponding sample denotations.

In the next step the AlO$_x$ layer is formed by static oxidation by flooding the deposition chamber with pure oxygen. Oxidation parameters like partial oxygen pressure, oxidation temperature or oxidation time were varied and sometimes plasma- or UV-enhanced oxidation was applied which leads to different AlO$_x$ thicknesses and O contents. We emphasize, that the study of the effect of the oxidation conditions on the oxygen concentration in the AlO$_x$-tunnel barrier is complex and will be presented in a separate publication. However, the variation of oxidation conditions does not affect the growth of the lower Al layer and the morphology of the Al surface at the Al/AlO$_x$ interface and can be neglected regarding conclusions about the Al growth of the lower Al layer. In this work we solely focus on the thickness homogeneity of the AlO$_x$ layers. In the last step, the upper Al layer is deposited using the same deposition parameters as for the lower Al layer.

Cross-section specimens for TEM were prepared by conventional mechanical preparation techniques as described by Strecker *et al.* [30] using Ar$^+$-ion milling with a Gatan 691 PIPS (*Gatan Inc., Pleasanton, USA*) as final preparation step. TEM and scanning transmission electron microscopy (STEM) were performed with a FEI Titan³ 80-300 (*Thermo Fisher Scientific, Waltham, USA*) operated at 300 kV. The instrument is equipped with an aberration corrector in the imaging lens system. Structure analyses were performed by comparing two-dimensional (2D) Fourier-transform patterns of high-resolution (HR)TEM images with simulated diffraction patterns using the *JEMS* software [31]. Bragg filtering is applied to visualize the behavior of selected lattice planes by selecting the corresponding reflections in the 2D Fourier-transform pattern with a digital aperture and subsequently performing an inverse 2D Fourier transformation.

The thickness of the $AlO_x$ layer was measured on the basis of HRTEM images by acquiring intensity line profiles with an integration width of 2 nm perpendicular to the $AlO_x$ layer. In such profiles the lattice planes of crystalline Al layers, in contrast to the amorphous $AlO_x$, show clear intensity maxima, and the distance between the uppermost lattice plane of the lower Al layer and the lowermost lattice plane of the upper Al layer can be measured.

## III. Experimental results and discussion

The results of our study are presented in three subsections. The first describes the optimization of the Si(111) substrate pretreatment. The second subsection focuses on the correlation of the deposition conditions (substrate temperature during Al deposition and Al-deposition rate) and structural properties of the lower Al layer, which determine the structural quality of the whole Al/$AlO_x$/Al-layer system. The analysis of the thickness homogeneity of the $AlO_x$ layer and the properties of the upper Al layer are presented in the third subsection.

### A. Pretreatment of the Si(111) substrate for optimization of the Al/Si(111) interface

Figure 1 shows HRTEM images of Al/Si(111) interfaces after different Si(111) surface treatments prior to the deposition of the lower Al layer. The protective resist layer was removed on all Si(111) substrates by a chemical cleaning procedure (cf. Experimental techniques). For the sample shown in Figure 1a, only plasma cleaning by the *kaufman source* in the deposition system was applied for further cleaning to remove remnant carbon contamination. After this process, the Si(111) substrate is still covered with a 3 nm thick native amorphous $SiO_x$ layer which obviously cannot be removed by plasma cleaning. Irrespective of the 3 nm SiO$_x$ layer, grain orientations in the deposited polycrystalline Al layer often do not deviate strongly from the [111] direction of the substrate as indicated in Figure 1a.

To fulfill the prerequisites for epitaxial Al growth, the SiO$_x$ layer must be completely removed and a 7x7 reconstructed Si(111) surface has to be prepared in the deposition system (see Experimental techniques). The success of the procedure is seen in the HRTEM image Figure 1b which demonstrates the complete lack of an amorphous SiO$_x$ layer and perfect alignment of the Al(111) planes parallel to the Si(111) substrate suggesting epitaxial growth of Al on Si(111). However, the Al layer does not grow as a single-crystalline layer, but forms Al grains, which can be rotated around the [111] direction or occasionally slightly tilted. In fact, grain size and grain orientation are strongly affected by the Al-deposition parameters and will be discussed in the next subsection. The large lattice-parameter mismatch of 25.5 % between Si and Al leads to the formation of dislocations at the Si(111)/Al(111) interface by the insertion of additional Al-lattice planes at the interface as shown in the Bragg-filtered HRTEM image in Figure 1c.

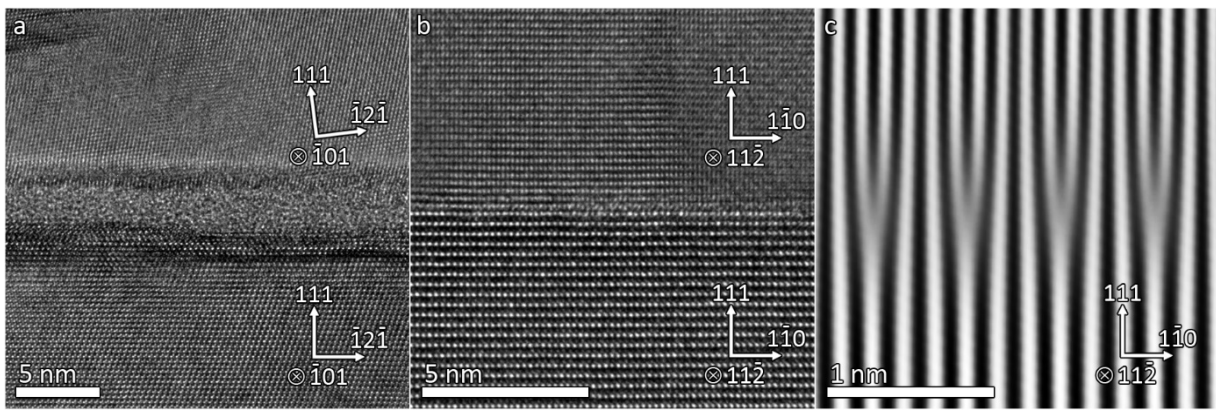

**Figure 1**. HRTEM images of the Al/Si(111) interface of Al/AlO$_x$/Al-layer systems with different substrate pretreatments. Chemical cleaning of the substrate to remove the protective resist was applied to all samples with (a) additional plasma cleaning in the deposition system or (b) additional HF-dip and high-temperature annealing at 700 °C for 20 min prior to Al deposition. (c) Magnified section of the Al(111)/Si(111) interface in (b) using Bragg filtering with the (1$\bar{1}$0)Al and (1$\bar{1}$0)Si planes.

We note that the SiO$_x$ layer could not be consistently removed by the HF-dip for all samples despite identical etching times. The etching rate of *BOE 7:1* was measured to be about 1 nm/s by a series of etching steps using 200 nm thick SiO$_2$ layers and with different etching times. A 45 s HF-dip should therefore have removed the SiO$_x$ layer completely. Thus, the Si surface must have been re-oxidized in some cases after the etching process by residual oxygen in the deposition system or even during transfer to the deposition system. Transfer time into the deposition chamber is therefore a critical parameter and should not exceed 1 min. Overall, HF-dip and a high-temperature heating step at 700 °C provides the best Si(111)/Al(111) interface that can be achieved in our HV deposition system.

**B. Dependence of the microstructure of the lower Al layer on deposition conditions**

To optimize the growth of the lower Al layer, five samples with different fabrication conditions regarding Al-deposition rate and substrate temperature were investigated. Si(111) substrates were subjected to a HF-dip and subsequent high-temperature treatment in all cases to obtain a clean and atomically flat Al(111)/Si(111) interface (cf. Figure 1b). Al deposition was performed under conditions listed in Table 1, i.e., at substrate temperatures between 300 °C and 100 °C with the same deposition rate (0.1 nm/s). Two further experiments were carried out at $T_s$=100 °C and increased deposition rates (0.5 nm/s and 1 nm/s).

The morphology of the lower Al layer is illustrated by overview cross-section bright-field STEM images of the complete Al/AlO$_x$/Al-layer systems in Figure 2 with the Si substrate, lower Al layer, AlO$_x$-tunnel barrier and upper Al layer. In the following we focus on the properties of the lower Al layer which are decisive for the structural quality of the whole layer system. For Al$_{300\_0.1}$ (Figure 2a), only large islands with varying lateral size and height are observed making such layers unsuitable for JJ fabrication. A continuous lower Al layer is formed at reduced $T_s$ for sample Al$_{200\_0.1}$. The homogeneity of the lower Al layer is further improved by reducing $T_s$ to 100 °C (Al$_{100\_0.1}$, Figure 2c) and increasing deposition rates (samples Al$_{100\_0.5}$ and Al$_{100\_1.0}$, Figures 2d,e). The homogenization of the structural properties is visualized by the homogenization of the thickness of the lower Al layer and the reduction of grain orientation variations, which can be recognized by different bright-field STEM intensities of grains related to Bragg-diffraction contrast. A homogeneous bright-field STEM intensity can be clearly recognized for samples Al$_{100\_0.5}$ and Al$_{100\_1.0}$ (Figures 2d,e) in contrast to Al$_{200\_0.1}$ and Al$_{100\_0.1}$ (Figures 2b,c) with a polycrystalline structure.

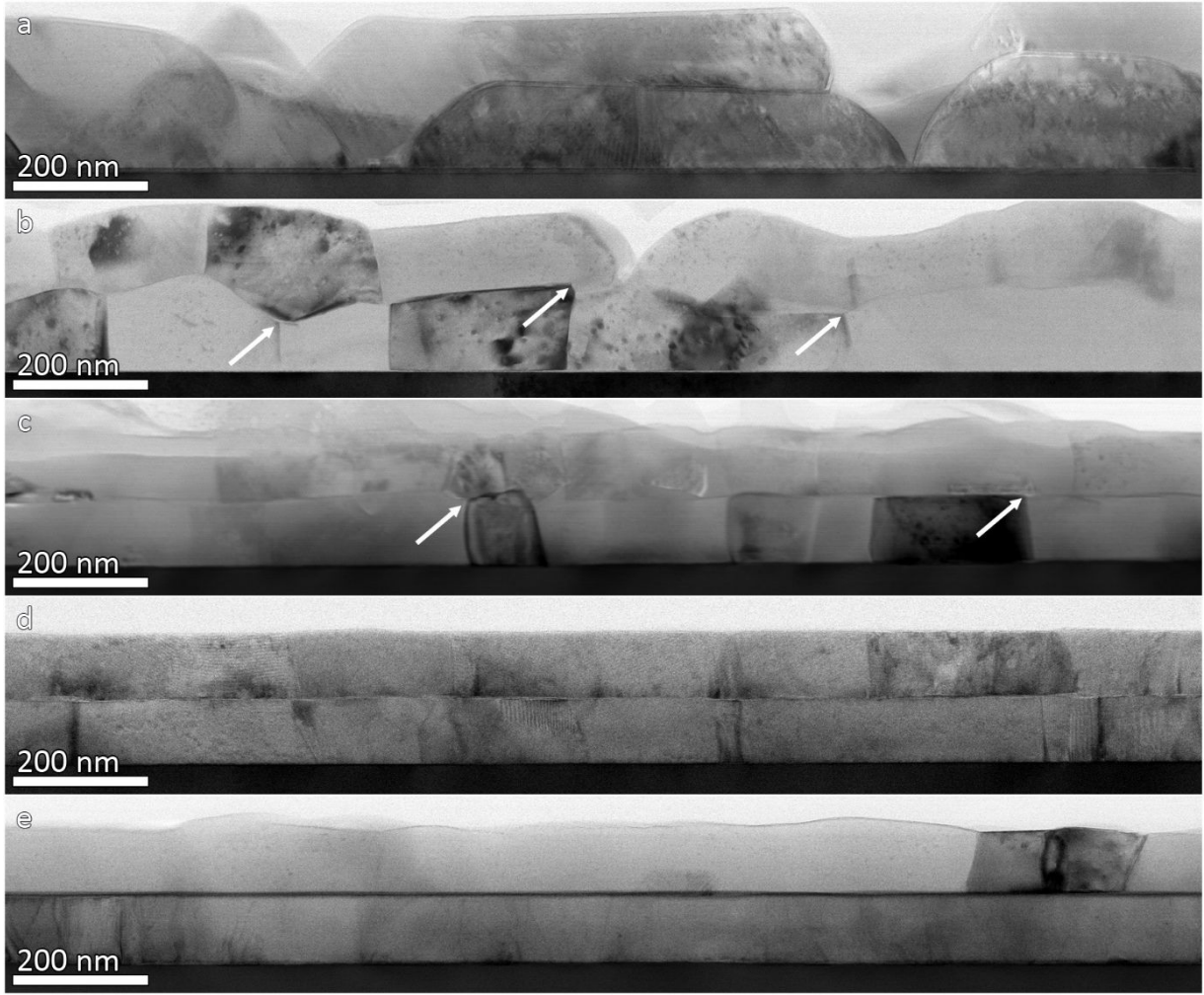

**Figure 2**. Bright-field STEM images of Al/AlO$_x$/Al-layer systems deposited at different substrate temperatures and Al-deposition rates on Si(111) substrates. (a) Al$_{300\_0.1}$, (b) Al$_{200\_0.1}$, (c) Al$_{100\_0.1}$, (d) Al$_{100\_0.5}$ and (e) Al$_{100\_1.0}$. The white arrows mark grain boundaries with pronounced grain-boundary grooving.

Measured lateral Al-grain sizes and Al-layer thicknesses illustrate the strong influence of $T_s$ and $r$ on the morphology of the lower Al layer. The roughness of Al layers is quantitatively determined by measuring grain thicknesses over lateral distances of 10 – 15 μm with one data point every 50 nm, which yields average values and standard deviations given in Table 2. The samples show a wide range of thickness variations $\Delta t$ from ±41.9 nm for Al$_{300\_0.1}$ due to island growth to the most homogeneous thickness for Al$_{100\_1.0}$ with $\Delta t$ of only ±1.6 nm. There is an obvious trend towards more homogeneous Al-layer thickness with decreasing $T_s$ and increasing $r$. The reasons for this behavior are visible in Figure 2. First, the grain surfaces flatten with decreasing $T_s$ and increasing $r$. The second effect that leads to thickness variations is grain boundary (GB) grooving which can also locally change the thickness of the AlO$_x$-tunnel barrier as discussed in detail by Nik *et al.* [8] and us [9]. The growth experiments in this work show that GB grooving depends strongly on $T_s$ and $r$ as demonstrated by Figure 2 where GB grooving is mainly observed in Al$_{200\_0.1}$ and Al$_{100\_0.1}$ (cf. arrows in Figures 2b,c). GB grooving is considerably reduced in Al$_{100\_0.5}$ (Figure 2d) and almost completely suppressed in Al$_{100\_1.0}$ (Figure 2e).

| sample | average lateral grain size [nm] | Al-layer thickness [nm] | AlO$_x$-layer thickness [nm] |
|---|---|---|---|
| Al$_{300\_0.1}$ | 375±116 | 114.1±41.9 | 1.62±0.29 |
| Al$_{200\_0.1}$ | 244±87 | 109.8±17.8 | 1.65±0.23 |
| Al$_{100\_0.1}$ | 200±71 | 99.0±6.2 | 1.73±0.19 |
| Al$_{100\_0.5}$ | 269±107 | 98.3±2.4 | 1.59±0.11 |
| Al$_{100\_1.0}$ | 347±208 | 98.8±1.6 | 4.88±0.17 |

**Table 2**. Average lateral grain size, thickness of the lower Al layer and thickness of the AlO$_x$ layer deposited at different temperatures and deposition rates.

The distribution of lateral grain sizes is presented in Figure 3 for all samples and average lateral grain sizes are given in Table 2. The average grain sizes decrease for a constant deposition rate of 0.1 nm/s with decreasing $T_s$. This trend is reversed if the deposition rate is increased for samples Al$_{100\_05}$ and Al$_{100\_1.0}$. The grain-size distribution can be well fitted by lognormal distributions for Al$_{200\_0.1}$, Al$_{100\_0.1}$ and Al$_{100\_0.5}$. Al$_{300\_0.1}$ does not show such a distribution which can be attributed to a different growth mode by the formation of large islands instead of a continuous Al layer. Another exception is sample Al$_{100\_1.0}$ which shows a large number of small grains between 50 nm and 150 nm and a broad range of grain sizes with lateral extensions up to 1 μm.

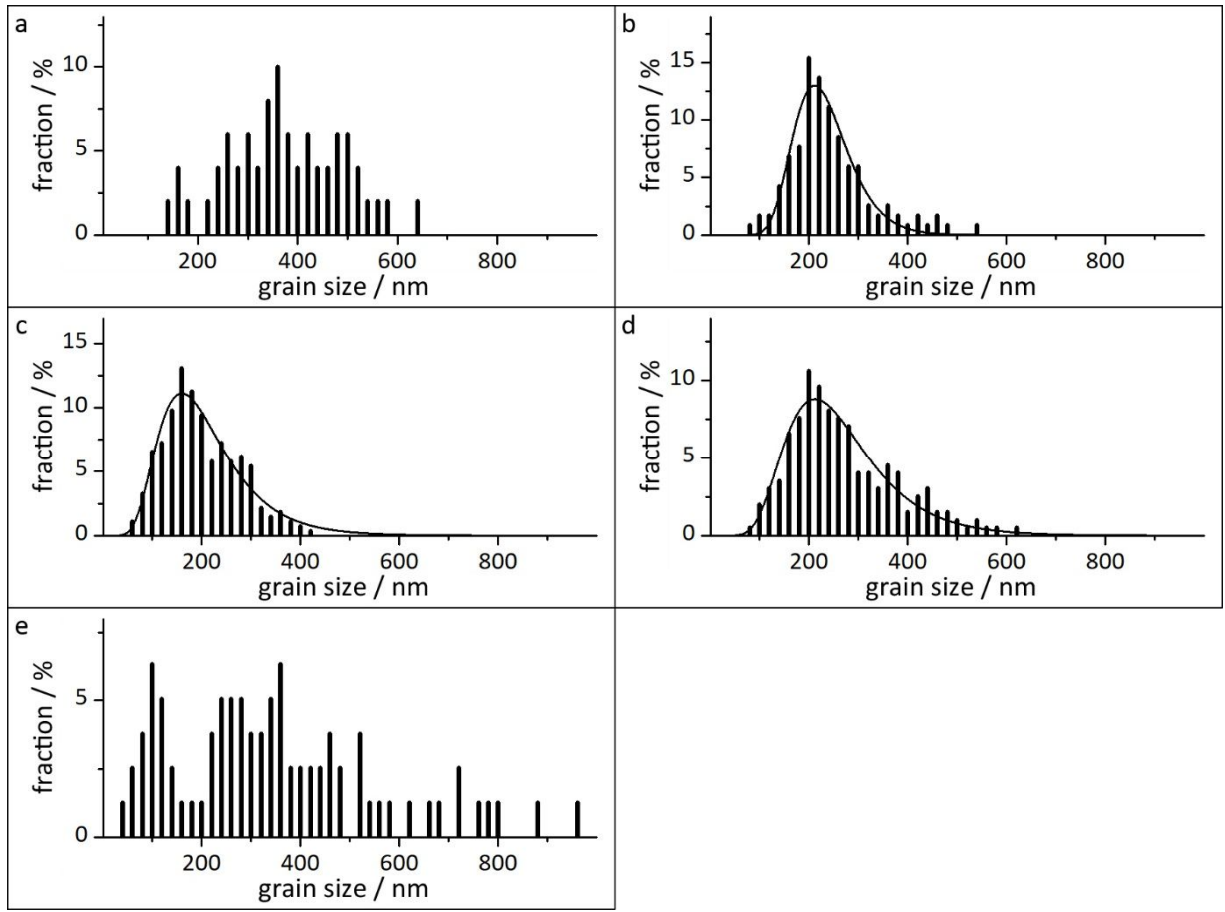

**Figure 3**. Distribution of lateral grain sizes in the lower Al layer for samples deposited at different substrate temperatures and deposition rates on Si(111). (a) $Al_{300\_0.1}$, (b) $Al_{200\_0.1}$, (c) $Al_{100\_0.1}$, (d) $Al_{100\_0.5}$ and (e) $Al_{100\_1.0}$ with fitted lognormal distributions (black line) in b, c, d.

The behavior of the average grain sizes and grain-size distributions can be understood by the following considerations. In the very initial deposition stage, Al islands are nucleated on the substrate which coalesce at some point to a closed film. The size of the islands at the stage of coalescence decreases and the number density increases with decreasing deposition temperature because Al-adatom mobility is reduced and the formation of large islands is prevented. After coalescence, grain coarsening occurs during further deposition to reduce the energy of the system that is stored in grain boundaries. Coarsening also depends on the grain-boundary mobility which is temperature dependent, i.e., coarsening is less pronounced at lower temperatures and leads to smaller (average) grain sizes as observed for $Al_{200\_0.1}$ and $Al_{100\_0.1}$ (Table 2). This coarsening behavior is denoted a normal grain growth [32–34] and is characterized by a lognormal grain-size distribution which is found for all samples apart from samples $Al_{300\_0.1}$, where complete coalescence of islands was not yet achieved, and $Al_{100\_1.0}$ (cf. Figure 3a,e). The reduction of average grain size is reversed for the samples that were grown at 100 °C with increased deposition rates. Higher deposition rates further impede Al-adatom mobility on the surface, i.e., the size of the islands at the stage of coalescence is further reduced. This increases the total grain boundary energy and leads to a larger driving force for grain coarsening. At high deposition rates and very small original island/grain sizes, the driving force for grain coarsening can be high enough that some grains grow to a huge size. This process is denoted as abnormal grain growth [35] and manifests itself by the failure to fit the grain-size distribution by a lognormal function (Figure 3e). This is clearly the case for sample $Al_{100\_1.0}$ where a wide distribution of grain sizes between less than 50 nm and 900 nm is observed. However, the increase of the driving force for grain coarsening is already observed for sample $Al_{100\_0.5}$ where the average grain size is already larger than for $Al_{100\_0.1}$. Abnormal grain growth may be additionally favored by the formation of large Al(111) surfaces which have the lowest surface energies of all Al surfaces [23] and are thus the preferred orientation for large grains [36].

In the following we analyze GBs in the lower Al layer in more detail for the layers deposited at 100 °C (cf. Figure 4) because this deposition temperature yields most homogenous Al layers in terms of layer thickness. Crystal orientations were determined by comparing 2D Fourier-transform patterns of HRTEM images with calculated diffraction patterns. This procedure allows to determine the orientation of the GB plane and the tilt angles between neighboring grains.

The HRTEM image Figure 4a shows a GB with pronounced GB grooving for a sample that was prepared under the same conditions as $Al_{100\_0.1}$, but with UV-enhanced oxidation leading to a thicker AlO$_x$ layer. The (111) lattice planes in the right grain are oriented parallel to the Si substrate while the left grain is not in epitaxial orientation resulting in a GB with low symmetry. The presumably high GB energy leads to strong GB grooving. Figure 4b shows a GB in Al$_{100\_0.1}$. The (111) planes of the two adjacent Al grains are almost parallel to the Si(111) substrate. Only a slight tilt by about 3° around the [$\bar{1}$01] direction is measured between the Al(111) planes in the two grains leading to a small-angle ($\bar{1}5\bar{1}$)/(1$\bar{4}$1) GB (we note that we determine the planes in the two adjacent grains that coincide at the GB). The GB is inclined with respect to the Al/AlO$_x$ interface. The small-angle ($\bar{1}5\bar{1}$)/(1$\bar{4}$1) GB does not induce significant GB grooving, but bending of the Al surface and a change of the crystallographic orientation of the Al surface at the Al/AlO$_x$ interface from Al(111) to Al(101) and back to Al(111). The AlO$_x$ layer on the Al(101) surface is ~5 % thicker than on the Al(111) surface which can be attributed to different oxidation rates on different Al surfaces. Other grains with higher-indexed Al surfaces at the Al/AlO$_x$ interface also show this effect like, e.g., a 10 – 15 % reduced AlO$_x$ thickness on an Al(131) surface compared to Al(111).

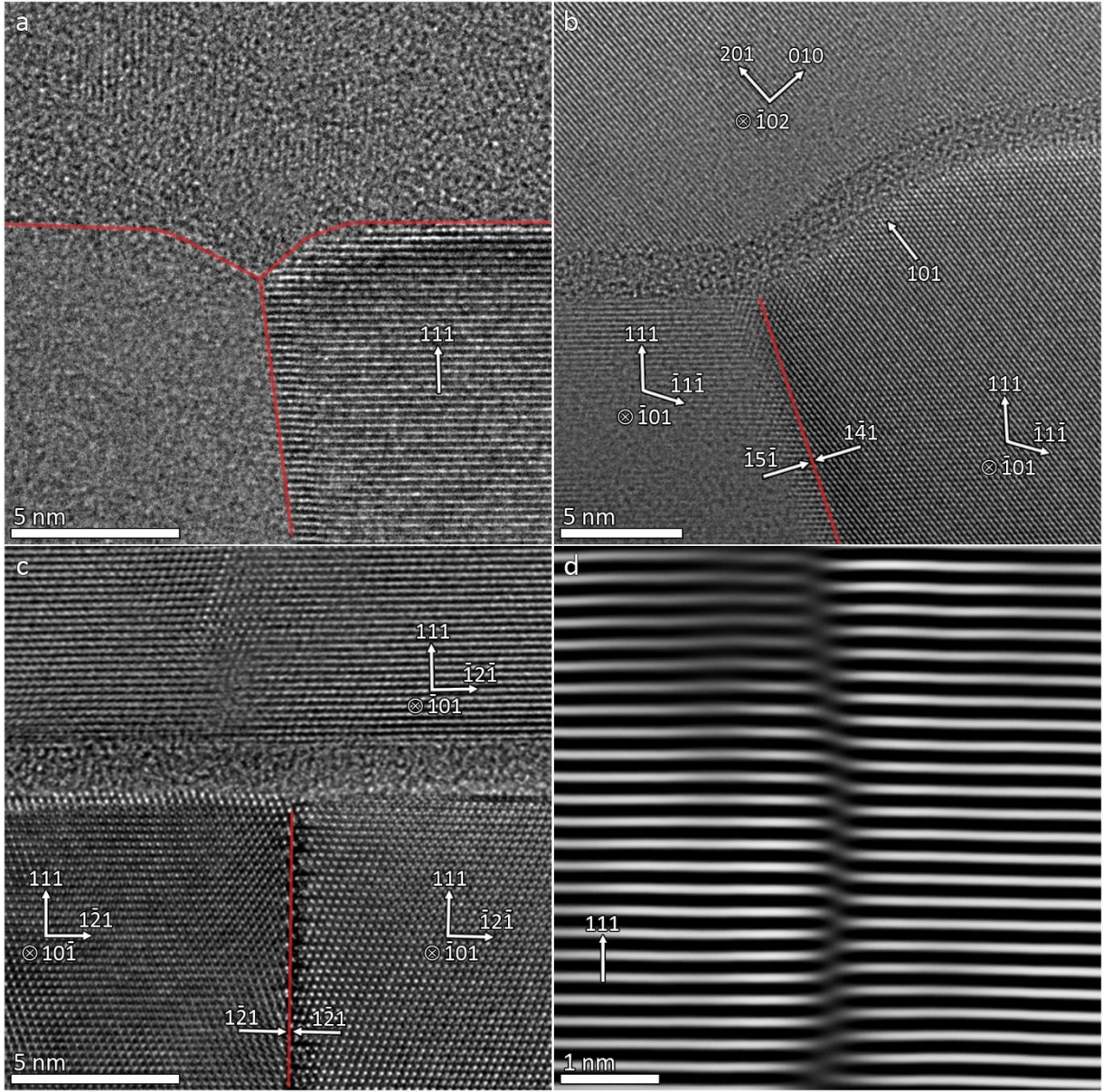

**Figure 4.** HRTEM images of GBs intersecting the Al/AlO$_x$ interface in (a) a sample that was grown under the same conditions as Al$_{100\_0.1}$ apart from the thicker AlO$_x$ layer where UV-enhanced oxidation was used, (b) Al$_{100\_0.1}$, (c) Al$_{100\_0.5}$ and (d) magnified section of the Al-grain boundary in (c) after Bragg filtering for Al(11$\bar{1}$) planes. The orientation of the Al grains in (c) is assumed to be rotated by 180 degrees around the [111] direction. The red lines delineate grain boundaries.

Besides the ($\bar{1}5\bar{1}$)/(1$\bar{4}$1) GB, a variety of different GB orientations such as ($\bar{1}\bar{1}\bar{1}$)/(1$\bar{1}$1), (101)/($\bar{1}$10) and others are observed in Al$_{100\_0.1}$. Figure 4c shows a typical GB in Al$_{100\_0.5}$ which is a symmetric tilt boundary of the type $\Sigma$=3/{112}, i.e., the number of coincidence lattice sites is three and the GB plane is a {112} plane in both grains which contains the <110> tilt axis. This twin boundary is a low-energy GB and occurs frequently in face-centered cubic metals [37–39]. Characteristic features are: (a) the (111) planes are parallel in the two neighboring grains, (b) the GB is oriented perpendicular to the Al/AlO$_x$ interface and (c) does not induces bending or grooving. Such a GB can be formed by a 180° rotation around the [111] direction. Al$_{100\_0.5}$ still contains other GBs which are formed if the Al(111) planes in neighboring grains are tilted against each other. This is the origin of less symmetrical GBs, which are in general not perpendicular to the Al/AlO$_x$ interface and cause bending or grooving like the GBs shown in Figures 4a,b. In contrast, we observed exclusively $\sum=3/\{112\}$ twin boundaries in Al$_{100\_1.0}$ leading to a highly planar Al/AlO$_x$ interface (cf. Figure 2e).

A closer look on the atomic structure of the Al layer and Al/Si interface explains the high content of $\sum=3/\{112\}$ GBs along <110> directions. It is well known that the Si(111) 7x7 surface contains steps parallel to the <101> directions which separate atomically flat terraces [40, 41]. A $\sum=3/\{112\}$ GB will be formed if grains on neighboring terraces are rotated by 180° around the [111] direction and reach a step. HRTEM images of Al$_{100\_1.0}$ show GBs that are solely oriented along <110> directions and thus support the hypothesis presented above. The steps at the Si(111) surface also lead to a small vertical displacement of the Al(111) lattice planes across the GB as shown in the Bragg-filtered HRTEM image (Figure 4d) where only the Al(111) planes are visible. $\sum=3/\{112\}$ GBs do not show measurable GB grooving due to the low GB energy [38] and low surface energy of the Al(111) planes [23].

It is on first sight surprising that only $\sum=3/\{112\}$ GBs are formed in Al$_{100\_1.0}$ whereas various GB types occur in all other samples. We attribute this effect to a change of the Al-growth mode. Lognormal grain-size distributions for Al$_{200\_0.1}$, Al$_{100\_0.1}$ and Al$_{100\_0.5}$ indicate normal grain-growth behavior leading to lateral average grain sizes 2 to 4 times of the film thickness [34] (cf. Table 2) while abnormal grain coarsening occurs for Al$_{100\_1.0}$. However, more studies are necessary to clarify GB formation in detail.

Overall, Al$_{100\_1.0}$ with its homogeneous epitaxial lower Al layer provides best conditions for the formation of an AlO$_x$ layer with constant thickness. Thus, the combined effects of grain properties and influence of GBs are decisive for the optimization of the structural properties of the lower Al layer. The benefit of comparatively large grain sizes (lower GB density) at high substrate temperatures is impaired by corrugated grain surfaces and high-energy GBs, which induce bending and grooving. An Al layer with optimum properties was fabricated at the lowest deposition temperature $T_s = 100\,°C$ and highest deposition rate $r = 1.0$ nm/s. We could not further reduce $T_s$ due to the lack of active substrate cooling in our deposition system, which increases the cooling time to up to a few hours after the high-temperature substrate treatment. Within this time interval the Si surface can be re-oxidized by residual oxygen, resulting in a thin $SiO_x$ layer which is detrimental to achieving epitaxial Al growth. For constant low $T_s$, increasing deposition rates (which are limited to 1.0 nm/s in our deposition system) lead to larger average grain sizes by anomalous grain growth. Increasing deposition rates also favor preferential formation of low-energy $\Sigma=3/\{112\}$ GBs which do not induce grooving or bending at the $Al/AlO_x$ interface.

## C. Properties of $AlO_x$ and the upper Al layer

After Al deposition the surface was oxidized by static oxidation with pure $O_2$ to form an amorphous $AlO_x$-tunnel barrier with a thickness of $1.5 - 2.0$ nm. Although the oxidation conditions (oxidation times, oxidation temperature and $O_2$-partial pressures) were varied to obtain $AlO_x$ with different properties (to be presented separately), we will show in the following that the homogeneity of the $AlO_x$ layer depends to a large degree on the surface roughness of the lower Al layer. We note, that plasma-enhanced oxidation was applied for $Al_{100\_1.0}$ to increase the oxygen content of the $AlO_x$ layer. This also lead to an increased $AlO_x$-layer thickness, making this layer unsuitable for the fabrication of Josephson junctions, but it provides useful information concerning the crystallographic orientation of the upper Al layer.

Also, the different AlO$_x$ properties of Al$_{100\_1.0}$ do not affect the conclusions regarding the optimization of the Al deposition in section III B.

Average values and standard deviations of the AlO$_x$-layer thickness were measured for all samples according to the procedure described in section II and are listed in Table 2. All layers have overall thicknesses between 1.59 nm and 1.73 nm apart from 4.88 nm for Al$_{100\_1.0}$ where plasma-enhanced oxidation was applied. The thickness variation of the AlO$_x$-layers $\Delta t$ improves with decreasing $T_s$ and increasing $r$ from 0.29 nm for Al$_{300\_0.1}$ to 0.11 nm for Al$_{100\_0.5}$ and shows the same trend as the thickness variation of the lower Al layer (cf. Table 2). On first sight, the AlO$_x$-layer thickness of Al$_{100\_1.0}$ with a slightly larger $\Delta t$ of 0.17 nm does not seem to follow the trend, but the overall thickness of the AlO$_x$ layer is by a factor of three larger due to the application of a plasma-enhanced oxidation process in this particular case. Although the absolute $\Delta t$ value of Al$_{100\_1.0}$ increases slightly with respect to Al$_{100\_0.5}$, the percentage of the thickness variation is reduced from 6.9 % for Al$_{100\_0.5}$ to 3.5 % for Al$_{100\_1.0}$. The reduction of $\Delta t$ can be in general attributed to the decrease of the content of high-energy GBs (and higher content of low-energy $\Sigma=3/\{112\}$ GBs) which reduces GB grooving and leads to a smoother Al/AlO$_x$ interface.

The small absolute $\Delta t$ increase in Al$_{100\_1.0}$ can be mainly attributed to the upper Al/AlO$_x$ interface, which is more corrugated due to random grain orientations in the upper Al layer on the comparatively thick AlO$_x$ layer (cf. Figure 5a). In contrast, an abrupt upper Al/AlO$_x$ interface is observed in Al$_{100\_0.5}$ with an epitaxial Al-grain orientation in the upper layer that is preserved across the thin AlO$_x$ layer (cf. Figure 5b). The HRTEM images in Figure 5 also demonstrate that, besides thickness variations due to GB grooving (cf. Figure 4a) or Al surfaces with different crystallographic orientation (cf. Figure 4b), thickness variations are caused by atomic steps at Al/AlO$_x$ interfaces even in epitaxial Al layers. A HRTEM image of Al$_{100\_0.5}$ (Figure 5b) shows such steps at the lower and upper Al/AlO$_x$ interface. The upper Al layer is tilted by about 0.5° from the [111] direction which leads to an increase of atomic steps. The thickness of the AlO$_x$ layer will inevitably vary if the steps do not occur at the same lateral position or the step density is not identical at the lower and upper Al/AlO$_x$ interfaces.

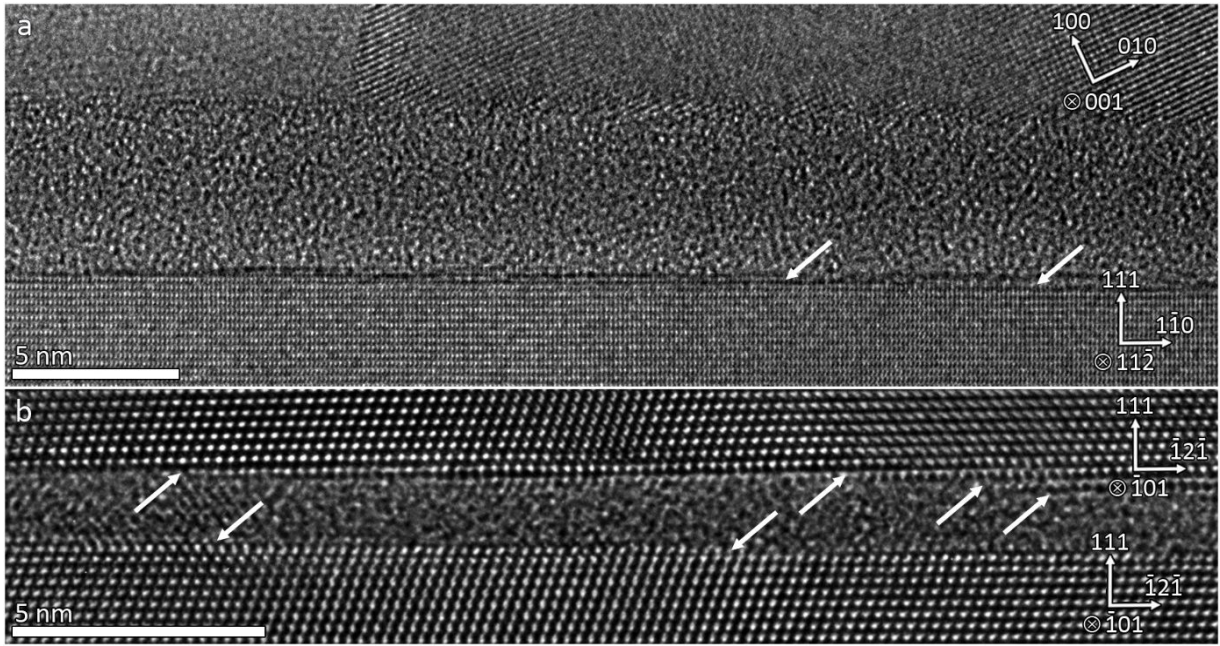

**Figure 5**. HRTEM images of Al/AlO$_x$ interfaces in (a) Al$_{100\_1.0}$ and (b) Al$_{100\_0.5}$. Atomic steps at the Al/AlO$_x$ interfaces are marked by white arrows.

These observations confirm once more that thickness variations of the lower Al layer correlate with thickness variations of the AlO$_x$ layer and emphasize the influence of structural properties of the lower Al layer on the AlO$_x$ layer. The HRTEM images in Figure 5 also visualize that the crystallographic orientation of grains in the upper Al layer also influences the thickness homogeneity of the AlO$_x$ layer. The thickness homogeneity of the AlO$_x$ layer can be optimized if the crystallographic orientation of lower and upper layer is identical. We note that the lower and upper Al layers were always deposited with the same deposition rate and nominally the same $T_s$ but deviations of about ±20 °C may have occurred for the upper layers depending on the oxidation temperature.

It is in general expected that grain orientations in the lower and upper Al layers are different due to the presence of the amorphous AlO$_x$ layer in between. This expectation is confirmed for Al$_{300\_0.1}$, Al$_{200\_0.1}$ and Al$_{100\_0.1}$ where grains in the upper Al layer show different Bragg contrast than grains in the lower Al layer (cf. Figures 2a,b,c). However, the upper Al layer in Al$_{100\_0.1}$ also contains some grains that grow with Al(111) lattice planes parallel to the lower Al layer, i.e., the information on the crystallographic orientation is transferred across the amorphous AlO$_x$ layer. Other Al grains in the upper layer are often only slightly misoriented with respect to Al(111). For Al$_{100\_0.5}$ the fraction of well aligned grains with Al(111) lattice planes parallel to the AlO$_x$ layer increases and only small orientation deviations between upper and lower Al grains are typically observed (cf. Figure 4c and Figure 5b). More random grain orientation are observed on a thicker AlO$_x$ layer as for Al$_{100\_1.0}$ with an AlO$_x$ thickness of 4.88 nm (cf. Figure 5a). However it is not only the AlO$_x$-layer thickness that determines the transfer of the orientation information, because the average AlO$_x$-layer thickness is almost identical for Al$_{300\_0.1}$, Al$_{200\_0.1}$, Al$_{100\_0.1}$ and Al$_{100\_0.5}$ (cf. Table 2). The phenomenon is only found in Al$_{100\_0.5}$ and to a lesser degree in Al$_{100\_0.1}$ where the Al-deposition conditions favor epitaxial growth. For the other samples, the deposition parameters lead to a larger variation of grain orientations in the lower Al layer despite a clean Si(111) surface, and it is reasonable that this behavior is pertained in the upper Al layer.

The transfer of crystallographic orientation from the lower to the upper Al layer was also found in molecular dynamics simulations by DuBois *et al.* [42], where Al grown on a thin amorphous AlO$_x$ layer (1.2 nm thick) tends to pick up the orientation of the lower Al layer. A possible explanation could be pinholes in the AlO$_x$ layer which form during the first stage of the oxidation [43]. However, this should only happen for ultra-thin AlO$_x$ layers (less than 0.6 – 0.8 nm), whereas thicker layers should form a continuous layer. Electrical measurements on JJs also showed a reduced tunneling resistance and increased leakage currents for a critical AlO$_x$-layer thickness below 1 nm [44, 45]. According to these observations, we can expect continuous AlO$_x$ layers in our samples because the critical thickness for pinhole formation is exceeded and HRTEM images do not indicate pinhole formation. We speculate that the periodic potential of the Al(111) surface across a thin AlO$_x$ layer can still be strong enough to initiate Al growth with the same orientation. Thus, despite of optimum Al-growth conditions, Al$_{100\_1.0}$ shows more random grain orientation than Al$_{100\_0.5}$ due to the increased AlO$_x$-layer thickness of 4.88 nm. In summary, epitaxial growth conditions (low temperatures and high deposition rates) combined with a thin AlO$_x$ layer with a thickness below 2 nm will lead to a well oriented upper Al layer and an AlO$_x$ layer with minimal thickness variations, but the origin and conditions of transfer of information on the crystallographic orientation across thin amorphous AlO$_x$ layers has to be further investigated.

## IV. Conclusions

Al/AlO$_x$/Al-layer systems for application in Josephson junctions were deposited on Si(111) substrates in a standard high-vacuum electron-beam deposition system in this work. It is demonstrated that optimization of the growth of the lower Al layer leads to epitaxial lower Al layers with the desired homogenization of the AlO$_x$ layer thickness and, correspondingly, an optimization of the properties of the whole Al/AlO$_x$/Al-layer system. The following conclusions can be drawn from the correlation of deposition parameters and the structural properties of the Al/AlO$_x$/Al-layer systems:

- HF-cleaning and high temperature treatment (in our case 700 °C for 20 min) are mandatory to achieve epitaxial growth of Al(111) on Si(111).

- Epitaxial growth of the lower Al layer on Si(111) was achieved for low substrate temperatures and high deposition rates (100 °C and 1 nm/s in our case). Under these conditions, grains with large lateral sizes and planar surfaces are formed due to abnormal grain growth. In addition, grooving and bending of the Al/AlO$_x$ interface and corresponding AlO$_x$ thickness variations are avoided because only $\sum=3/\{112\}$ symmetrical twin boundaries occur in the lower Al layer. Elimination of other GB types, which are formed at higher substrate temperatures and lower deposition rates, improves the planarity of the Al/AlO$_x$ interface and homogenizes the AlO$_x$-layer thickness. It is also favorable that under these conditions the information of the crystallographic orientation of the lower Al layer is transferred across AlO$_x$ layers with a thickness below 2 nm.

Further reduction of the substrate temperature during Al deposition may be beneficial because a transition from a growth mode, which is dominated by grain nucleation and grain growth, to two-dimensional layer-by-layer growth may occur. This requires active substrate cooling to keep the time between high-temperature Si-substrate treatment and start of the Al deposition as short as possible to avoid re-oxidation and contamination of the Si substrate. We also point out that our study may pave the way to grow crystalline Al$_2$O$_3$-tunnel barriers on epitaxial Al(111) which has been achieved up to now only in an UHV system and which may be beneficial for reducing noise in Josephson junction-based superconducting devices.

**Acknowledgements**

We thank Dr. Silvia Diewald (CFN Nanostructure Service Laboratory, Karlsruhe Institute of Technology) for carrying out the HF dip and Dr. Johannes Rotzinger (Physikalisches Institut, Karlsruhe Institute of Technology) for enlightening discussions.